\documentclass{emulateapj}
\usepackage{times}
\usepackage{wrapfig}
\usepackage{amsmath}
\usepackage{graphicx, subfigure}
\begin{document}
\title{Variable Hard X-ray Emission from the Candidate Accreting Black Hole in\\ Dwarf Galaxy Henize 2--10}

\slugcomment{Accepted for publication in The Astrophysical Journal}

\newcommand{\xmm}{{\it XMM}}
\newcommand{\xmmnewton}{{\it XMM-Newton}}
\newcommand{\chandra}{{\it Chandra}}
\newcommand{\asca}{{\it ASCA}}
\newcommand{\msun}{\ensuremath{M_{\sun}}}

\shorttitle{HENIZE 2-10 AGN}
\shortauthors{WHALEN}
\author{Thomas J. Whalen\altaffilmark{1,2}}
\author{Ryan C.\ Hickox\altaffilmark{1}}
\author{Amy E.\ Reines\altaffilmark{3}$^{,}$\altaffilmark{4}}
\author{Jenny E.\ Greene\altaffilmark{5}}
\author{Gregory R.\ Sivakoff\altaffilmark{6}}
\author{Kelsey E.\ Johnson\altaffilmark{7}}
\author{David M.\ Alexander\altaffilmark{8}}
\author{Andy D.\ Goulding\altaffilmark{5}}

\altaffiltext{1}{Department of Physics and Astronomy, Dartmouth College, 6127 Wilder Laboratory, Hanover, NH 03755; ryan.c.hickox@dartmouth.edu.}
\altaffiltext{2}{Harvard-Smithsonian Center for Astrophysics, 60 Garden Street, Cambridge, MA  02138, USA}
\altaffiltext{3}{Department of Astronomy, University of Michigan, 1085
South University Ave., Ann Arbor, MI 48109, USA}
\altaffiltext{4}{Hubble Fellow}
\altaffiltext{5}{Department of Astrophysical Sciences, Princeton University, Princeton, NJ 08544, USA}
\altaffiltext{6}{Department of Physics, University of Alberta, Edmonton, Alberta T6G 2E1, Canada}
\altaffiltext{7}{Astronomy Department, University of Virginia, Charlottesville, VA 22904, USA}
\altaffiltext{8}{Department of Physics, Durham University, South Road, Durham, DH1 3LE, United Kingdom}

\begin{abstract}
We present an analysis of the X-ray spectrum and long-term variability
of the nearby dwarf starburst galaxy Henize 2--10. Recent observations
suggest that this galaxy hosts an actively accreting black hole with
mass \textasciitilde $10^6$ \msun. The presence of an AGN in a
low-mass starburst galaxy marks a new environment for active galactic
nuclei (AGNs), with implications for the processes by which ``seed''
black holes may form in the early Universe.  In this paper, we analyze
four epochs of X-ray observations of Henize 2--10, to characterize the
long-term behavior of its hard nuclear emission. We analyze
observations with {\em Chandra} from 2001 and {\em XMM-Newton} from
2004 and 2011, as well as an earlier, less sensitive observation with
{\em ASCA} from 1997. Based on detailed analysis of the source and
background, we find that the hard (2--10 keV) flux of the putative AGN
has decreased by approximately an order of magnitude between the 2001
\chandra\ observation and exposures with \xmmnewton\ in 2004 and
2011. The observed variability confirms that the emission is due to a
single source. It is unlikely that the variable flux is due to a
supernova or ultraluminous X-ray source, based on the observed
long-term behavior of the X-ray and radio emission, while the observed
X-ray variability is consistent with the behavior of well-studied AGNs.
\end{abstract}


\section{Introduction}
\label{sec:intro}

The nearby (\textasciitilde 9 Mpc) dwarf starburst galaxy Henize
2--10, exhibits intense star formation \cite[e.g.,][]{alle76}, while
in the center of the galaxy, an X-ray point source \citep{kobu10he2}
and relatively luminous radio point source \citep{kobu99, john03he2}
were found to be co-spatial, suggesting the existence of an accreting
low-mass active galactic nucleus (AGN) with black hole (BH) of mass
\textasciitilde $10^6 \msun$ \citep{rein11he2, rein12he2}. This
represented the first possible detection of an AGN in a dwarf starburst
galaxy. Even if a large fraction of dwarf galaxies host massive BHs, they are challenging to detect as AGNs because the AGN emission is faint
and its signatures can be swamped by surrounding star formation
\citep[e.g.,][]{rein13dwarf}; X-ray observations can be one of the
most effective methods for identifying AGN in dwarf galaxies
\citep[e.g.,][]{rein14mrk709,lemo15xdwarf,secr15xdwarf}.  If the existence of an
AGN in He 2--10 is confirmed, it would serve as one of the best
possible analogs for BH and galaxy growth in the early history of the
Universe \cite[e.g.,][]{rein11he2}. Most bulge-dominated galaxies
contain supermassive BHs, however, the process by which the orignal
``seed'' BHs formed remains poorly constrained
\citep[e.g.,][]{john07firstbh,volo10bhform}.

\begin{deluxetable*}{cccccc}
\tabletypesize{\small}
\tablecolumns{6}
\tablecaption{Observation Details \label{table:obs}}
\tablehead{& & & & & 	\colhead{Net counts}\\
	\colhead{Instrument} &
	\colhead{Detector} &
	\colhead{Obs.\ ID} &
	\colhead{Date} &
	\colhead{Exposure in ks (clean)} &
\colhead{0.5--2 keV (2--8 keV)}}
\startdata
\emph{Chandra}         & ACIS-S & 2075 & 2001-03-23 &   20.0 (19.7) & \phn 983 (174) \\
\emph{XMM-Newton} & pn       & 0202650101 & 2004-05-27 &  42.0 (29.3) & 3216 (234) \\
\emph{XMM-Newton} & pn & 0672800101 & 2011-05-11 & 26.9 (17.6) & 1863 \phn (92) \\
\emph{ASCA}             & SIS       & 65017000 & 1997-11-30 & 39.8 (22.4) & \phn 197 \phn (52) 
\enddata

\tablecomments{The details of each of the observations used in this
  paper. Observed net counts after background subtraction are
  listed for the 0.5--2 and 2--8 keV bands (see
  \S~\ref{sec:reduction} for details of source extraction and
  background analysis). Exposure times listed are for the total
  exposure and the net exposure after cleaning for flares. We focus on
  our analysis on the more sensitive {\em Chandra} and {\em XMM}
  observations, but include the earlier {\em ASCA} observation as a
  check on the baseline flux level for the source.}
\end{deluxetable*}

Currently, the available observational evidence for the central
compact sources in He 2--10 favors its interpretation as a
supermassive BH. The majority of its radio emission originates
from a region $< 3$ pc $ \times 1$ pc in size \citep{rein12he2} and is
consistent with being spatially coincident with the \chandra\ hard
X-ray point source at the dynamical center of the galaxy
\citep{rein11he2}.  Assuming that the radio and X-ray emission are
produced by a BH, a comparison with the BH
fundamental plane \citep{merl03bhplane} suggests that the mass is
$\sim$10$^6$ \msun\ \citep{rein11he2}. Alternatively, the X-ray
emission could in principle come from an ultraluminous X-ray source
that is powered by a stellar-mass BH \citep{robe07ulx}. However this
cannot account for the oberved compact radio flux
\citep[e.g.][]{midd13ulx, wolt14ulx}, although we note that previous
radio and X-ray observations are not strictly simultaneous. We can
most likely rule out supernova (SN) remnants as the cause of the X-ray
emission; there are no massive star-forming clusters coincident with
the compact radio emission, rendering this scenario somewhat
implausible, but not impossible. To more robustly constrain the nature
of the compact central source in He 2--10 and to better constrain its
mass, it is important to understand how its X-ray luminosity varies
with time. This is the goal of the present paper.

 The original evidence for the AGN in He 2--10 came in part from
 analysis of the spectrally hard, resolved point source in the 2001
 \chandra\ observations. Here, we analyze data taken from
 \chandra\ (2001), \xmmnewton\ (2004 and 2011), and \asca\ (1997) to
 obtain spectra at each epoch and a resulting measure of the long-term
 variability of the hard nuclear source.  The temporal baseline of the
 observations is sufficient to probe variability on timescales
 reasonable for an intermediate-mass BH or low-mass AGN, as
 shown by the small known sampling of these rare objects
 \citep{dewa08}.

\section{Data Reduction and Spectral Analysis}
\label{sec:reduction}

In this section we describe the spectral extraction process for each
observation, with a focus on details of the sophisticated background
modeling, which was required for the \xmmnewton\ and
\asca\ observations. All spectral analyses are performed using {XSPEC
  v12.8.0} \citep{arna96xsp}. See Table~\ref{table:obs} for the
details of the observations. We will focus primarily on the more
sensitive observations taken by {\em Chandra} in 2001 and {\em XMM} in
2004 and 2011, but will also include a discussion of the 1997
\asca\ observation that provides a (less sensitive) baseline
measurement of the source flux. Throughout the paper, uncertainties on
X-ray measurements (fluxes, luminosities, and spectral parameters)
represent 90\% confidence intervals.

\subsection{\chandra}

\label{sec:chandra}

\begin{figure*}
\epsscale{0.7}
\plotone{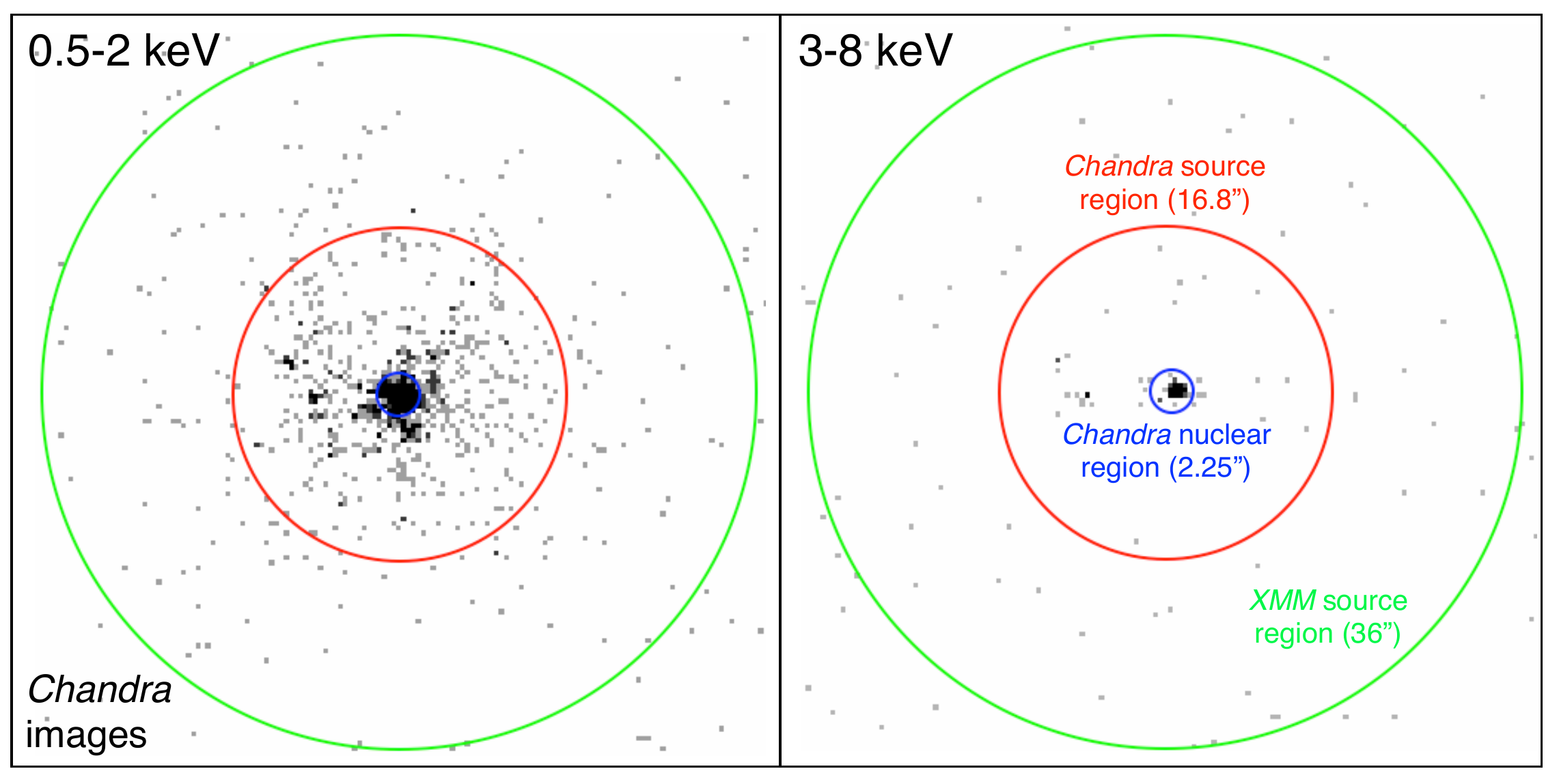}
\caption{\chandra\ images in the soft (0.5--2 keV) and hard (3--8 keV)
  X-rays; energy ranges are chosen to clearly separate the soft
  diffuse component from the hard compact nuclear emission. The
  \xmm\ source extraction region, as well as the two \chandra\ regions
  are superimposed. The excellent angular resolution of
  \chandra\ allows for clear imaging of the X-ray
  morphology. Widespread diffuse emission from star formation is seen
  in soft X-rays, while the central nuclear source is clearly seen at
  hard X-rays. \xmmnewton\ has significantly poorer angular
  resolution, thus a larger extraction region was required in order to
  include sufficient source flux.}
\label{fig:area}
\end{figure*}

The nuclear X-ray point source in He 2--10 was discovered in the 20 ks
ACIS-S \chandra\ observation of He 2--10 on 23 March 2001
\citep{rein11he2}. The pipeline-reduced data from this observation
(see Figure \ref{fig:area}) were obtained from the HEASARC public
archive and were reduced using the analysis tools of CIAO 4.5. Time
filtering yielded 19.7 ks of cleaned exposure. For \chandra\ we
defined two separate source extraction regions: a small region of
radius 2.25\arcsec\ at the nuclear region of the galaxy was used to
measure the flux of the hard point source, and a larger region of
radius 16.8\arcsec\ was used to include the soft diffuse X-ray
component. In either case, the background region was comprised of an
annulus of outer radius 58 arcsec, excluding the larger (diffuse)
source region and another point source about 45 arcsec away from the
source. We note that the background level above 2 keV in the {\em
  Chandra} observation was much less prominent than in the \xmm\ and
\asca\ observations.  The spectrum, as well as the response and
ancilliary files, were extracted using the {\tt specextract}
command. Due to the high signal-to-noise ratio and low background, we
used CIAO's built-in background subtraction rather than simultaneously
fitting source and background spectra, as for the \xmm\ and
\asca\ observations.

For all the observations presented in this paper, the source spectrum
is described with a model consisting of the power-law component to
model the hard nuclear source and optically-thin thermal (VMEKAL;
\citealt{mewe85mekal, lied95mekal}x) component to model the diffuse
emission, with abundance values fixed to those obtained in
\citet{kobu10he2} (0.78 for light elements and 0.29 for heavy
elements), and allowing the normalization and temperature to float.
We use a VMEKAL to match the spectral analysis of \citet{rein11he2},
but note that fitting with an APEC model \citep{smit01} has no
significant effect on the results for the hard component. Following
\citet{kobu10he2}, we include Galactic absorption $N_{\rm H,Gal} =
5\times10^{20}$ cm$^{-2}$ on all components, and local absorption for
the VMEKAL component ($N_{\rm H,Diffuse} = 9.7\times10^{20}$
cm$^{-2}$). The absorption on the power law component ($N_{\rm
  H,Nuclear}$) was allowed to float. Absorption is computed using the
tbabs model \citep{wilm00tbabs}. In XSPEC notation, the source model
is given by:

\begin{equation}
{\tt Source} = \textrm{tbabs$_{\rm Gal}$(tbabs$_{\rm Diffuse}$*VMEKAL + tbabs$_{\rm Nuclear}$*powerlaw)}
\label{eqn:source}
\end{equation}

We first fit this model to the spectrum from the nuclear
(2.25\arcsec\ radius) source region, to obtain the strongest possible
constraint on the emission from the unresolved hard
component. With relatively few counts at high energies we obtain poor
constraints on the hard X-ray photon index, so this is fixed to the
canonical AGN value of $\Gamma=1.8$. (Repeating the fit for values of
intrinsic $\Gamma$ varying within a range typical for AGN,
$1.4<\Gamma<2.2$ \citep{tozz06}, produces no significant change in the
unabsorbed flux.) The fit yields $N_{\rm H} = (4.61^{+1.67}_{-1.26})
\times 10^{22}$ cm$^{-2}$, indicating substantial absorption. We next
fit the same model to the spectrum from the extended source region (a
circle centered on the nuclear region with 16.8\arcsec\ radius), but
fix the $N_{\rm H}$ value on the power law component to that obtained for the nuclear spectrum. We
obtain a consistent and nearly identical flux for the hard power-law
component between the extended and nuclear regions, but a
substantially brighter diffuse (VMEKAL) component in the extended
region, with best-fit $k_T\ = 0.65\pm0.03$ keV. This confirms that for
an even larger extraction region such as those used for \xmm\ and
\asca, the hard spectral component can be associated with the compact
nuclear source.  The best-fit fluxes and spectral parameters are given
in Table \ref{table:params}.  We quote intrinsic (unabsorbed) fluxes
and luminosities for the hard nuclear component, and observed
(absorbed) fluxes for the soft diffuse component. For direct
comparison with the {\em XMM} analysis, we have also extracted a {\em
  Chandra} spectrum with somewhat larger radius of 36\arcsec,
corresponding to the {\em XMM} source region described in the next
section (Figure~\ref{fig:area}). Using this larger source region has
no significant effect on the spectral parameters.

\subsection{XMM-Newton}

\label{sec:xmm}

  We use two subsequent observations of He 2--10 by \xmm\ to
   constrain the long-term variability of the source. The
   \xmm\ observations on 27 May 2004 and 11 May 2011 have exposure
   times of 42 and 27 ks, respectively. For both \xmm\ observations,
 we reduced, cleaned, and extracted spectra from all three CCD
 cameras: pn, MOS1, and MOS2. After spectral extraction and analysis,
 the MOS1 and MOS2 data yielded significantly lower signal-to-noise
 ratios at energies $>$2 keV compared to the pn detector, so that no
 useful constraints were obtained on the hard emission from the
 nuclear point source. In what follows we therefore focus on results
 from the pn. 

The source extraction regions were 36\arcsec\ in radius, chosen to
provide a balance between extracting as many counts as possible from
the source and minimizing the background. (As a check, we have repeated the
analysis using a smaller source region of 25\arcsec\ radius, and obtain
essentially identical results with marginally larger uncertainties.)
The \chandra\ images (Figure~\ref{fig:area}) show that the extent of
both the nuclear and diffuse components are substantially smaller than
the 36\arcsec\ extraction region, such that they can both be considered
as approximate point sources for the \xmm\ analysis. In both data sets
the source region was on-axis and did not lie on any chip gaps. We
extracted pn spectra for counts in the energy range 0.2--15 keV and
with event patterns 0--4. Response files were produced from the
\xmmnewton\ Current Calibration Files corresponding to the time of
this observation. The source ARF is calculated including a correction
for photons falling outside the extraction region. This
energy-encircled fraction (EEF) varies with energy but is
$\approx$85\% at 5 keV.  The source spectrum is described with the
same model as for {\em Chandra}, consisting of VMEKAL and power law
components (Equation~\ref{eqn:source}).

To maximize the number of counts in the background spectra and thus
achieve the highest possible signal-to-noise ratio, we extracted the
background spectrum from a large annulus of outer radius 3 arcmin
around the source. Based on a number of trials, the 3 arcmin annulus
was determined to provide the optimum number of background counts
without needing to account for the variation in background flux at
larger off-axis angles. Using the SAS command {\tt edetect\_chain}, 5
sources in the pn field of view were detected and subsequently
excluded from the background region. With these sources excluded, the
background region is $\approx$23 arcmin$^2$ in area, or 20 times
larger than the source region. Because the background emission is
extended in nature, the ARF for the background region did not include
an EEF correction.

The pn background spectrum was fitted with two components.  The
instrumental background is modeled by a power law continuum, plus
Gaussian emission lines caused by fluorescence (Al K-$\alpha$ at 1.5
keV and CuNi K-$\alpha$ at 8.5 keV \citep{cart07xmm}. Because this
instrumental background is produced internally to the detector and is
not affected by the mirror response, in modeling the observed counts
it was was convolved with an RMF but not multiplied by an ARF. The
background line energies were determined from fitting each line
individually, and then fixed for the full spectral analysis, while the
intrinsic line widths are fixed to be consistent with zero.

The sky background component is dominated by the diffuse soft cosmic
X-ray background (CXB), which can be modeled as thin-thermal emission
\citep{hick06a, hick07b}. We used an APEC model for this component, to
match the spectral shape obtained by \citet{hick06a} in fitting the
unresolved CXB spectrum in the \chandra\ Deep Fields, fixing $kT=0.17$
keV and allowing the normalization to float. The hard ($>2$ keV) CXB
can be described as a power law owing to the summed emission from a
large number of AGN \citep[e.g.,][]{hick06a}. We did not include this
as a separate component here, as the average expected numbers of
counts is $<$2\% of the instrumental background, so its small
contribution to the total background flux can be effectively accounted
for in our modeling of the instrumental background.

The full model for the total observed emission in the \xmm\ source
region is:

\begin{equation}
\label{eqn:xmm}
{\tt Data} = {\tt Source} + {\tt Instrumental\; BG} + {\tt Sky\; BG}.
\end{equation}

The source spectrum is modeled by an absorbed VMEKAL and power law (Equation~\ref{eqn:source}), while the background components are modeled, in XSPEC notation, as:

\begin{equation}
\label{eqn:instbg}
{\tt Instrumental\;BG} = \textrm{powerlaw + gauss + gauss + gauss} \\
\end{equation}
and
\begin{equation}
\label{eqn:skybg}
{\tt Sky\; BG} = \textrm{APEC},
\end{equation}
where ${\tt Instrumental\; BG}$ is convolved with the RMF only, while ${\tt Source}$ and ${\tt Sky\; BG}$ are convolved by the RMF and multiplied by the ARF.

\begin{figure*}
\epsscale{0.7}
\subfigure{
	\includegraphics[width=0.5\textwidth]{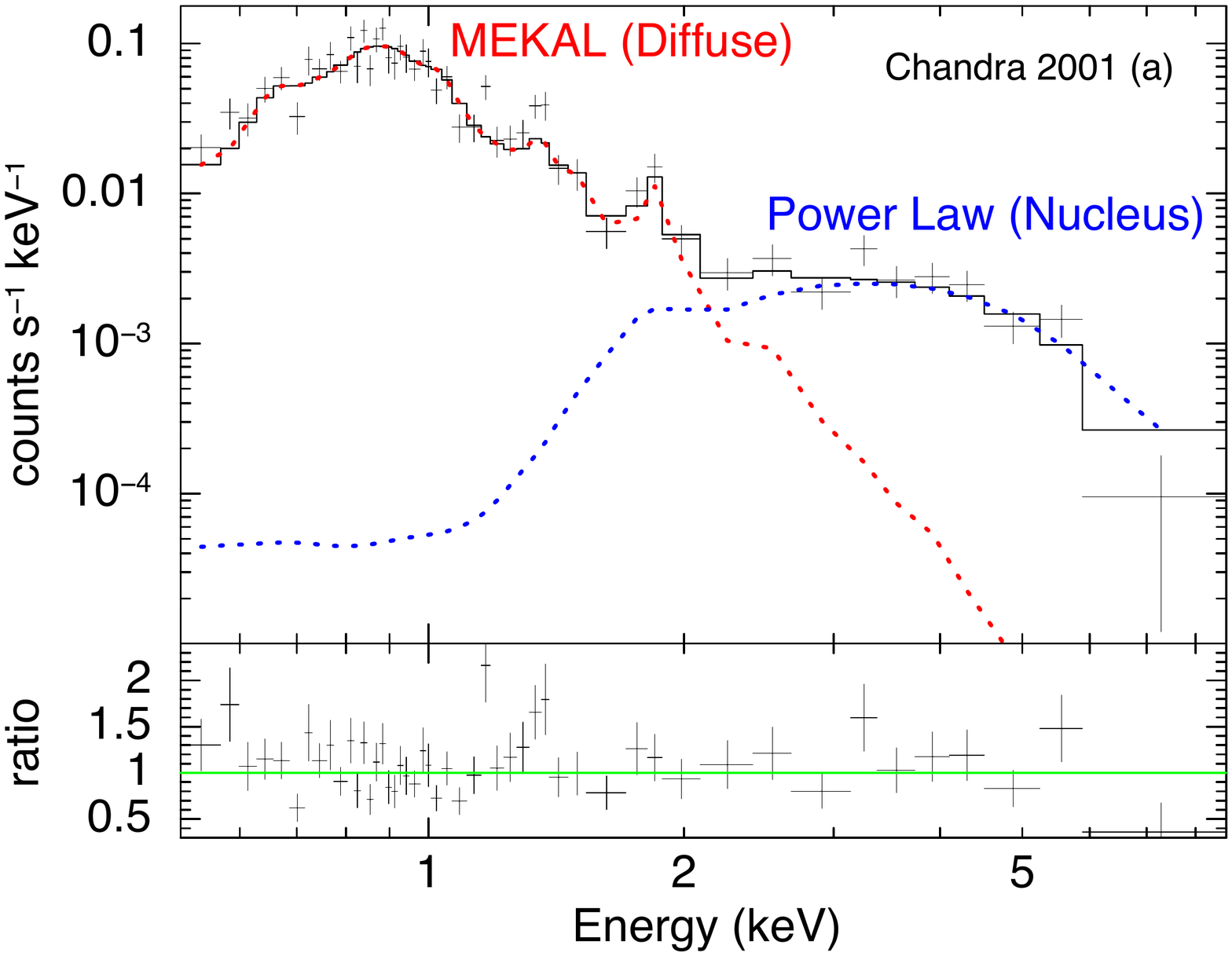}}
\subfigure{
	\includegraphics[width=0.5\textwidth]{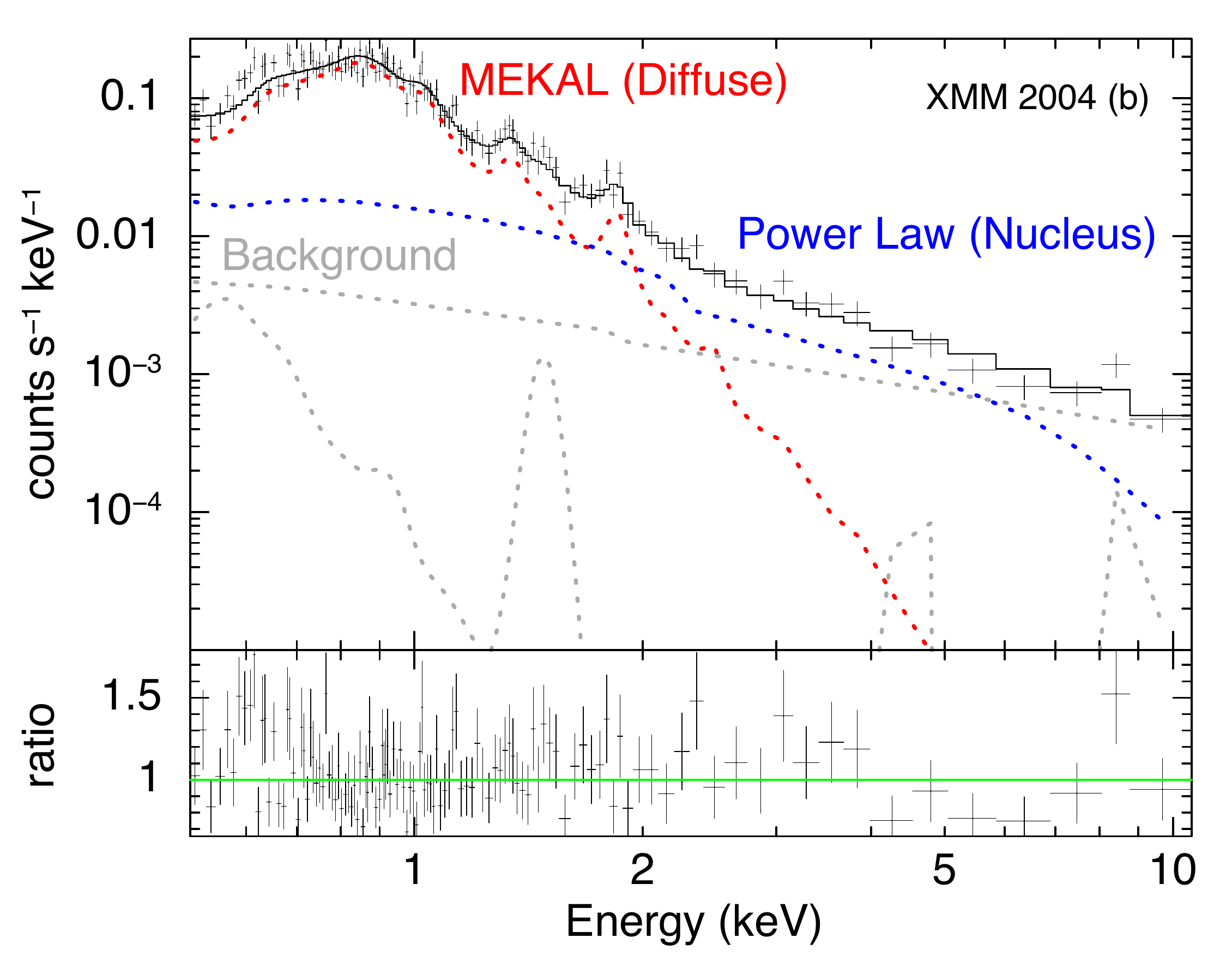}}
\subfigure{
	\includegraphics[width=0.5\textwidth]{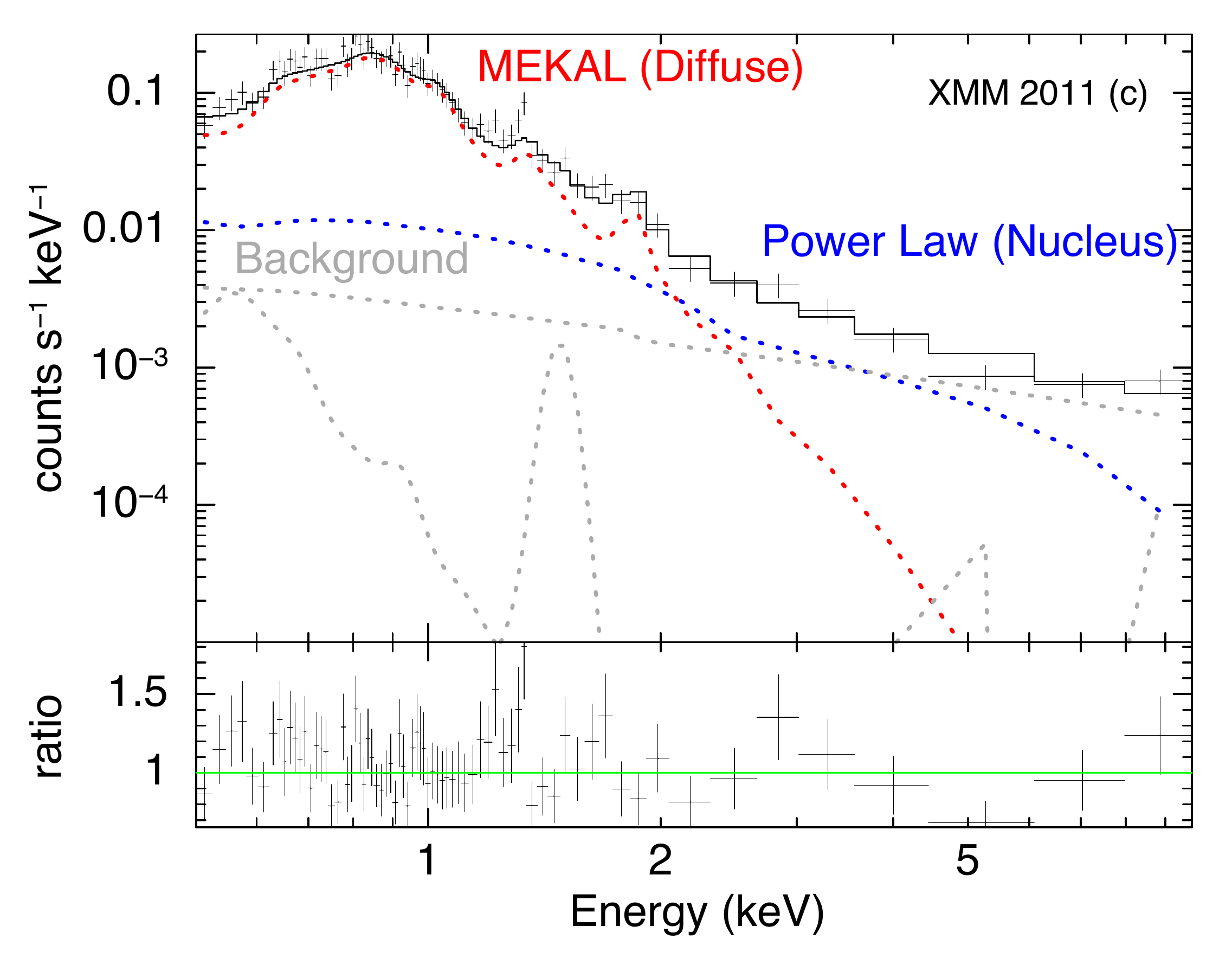}}
\subfigure{
	\includegraphics[width=0.5\textwidth]{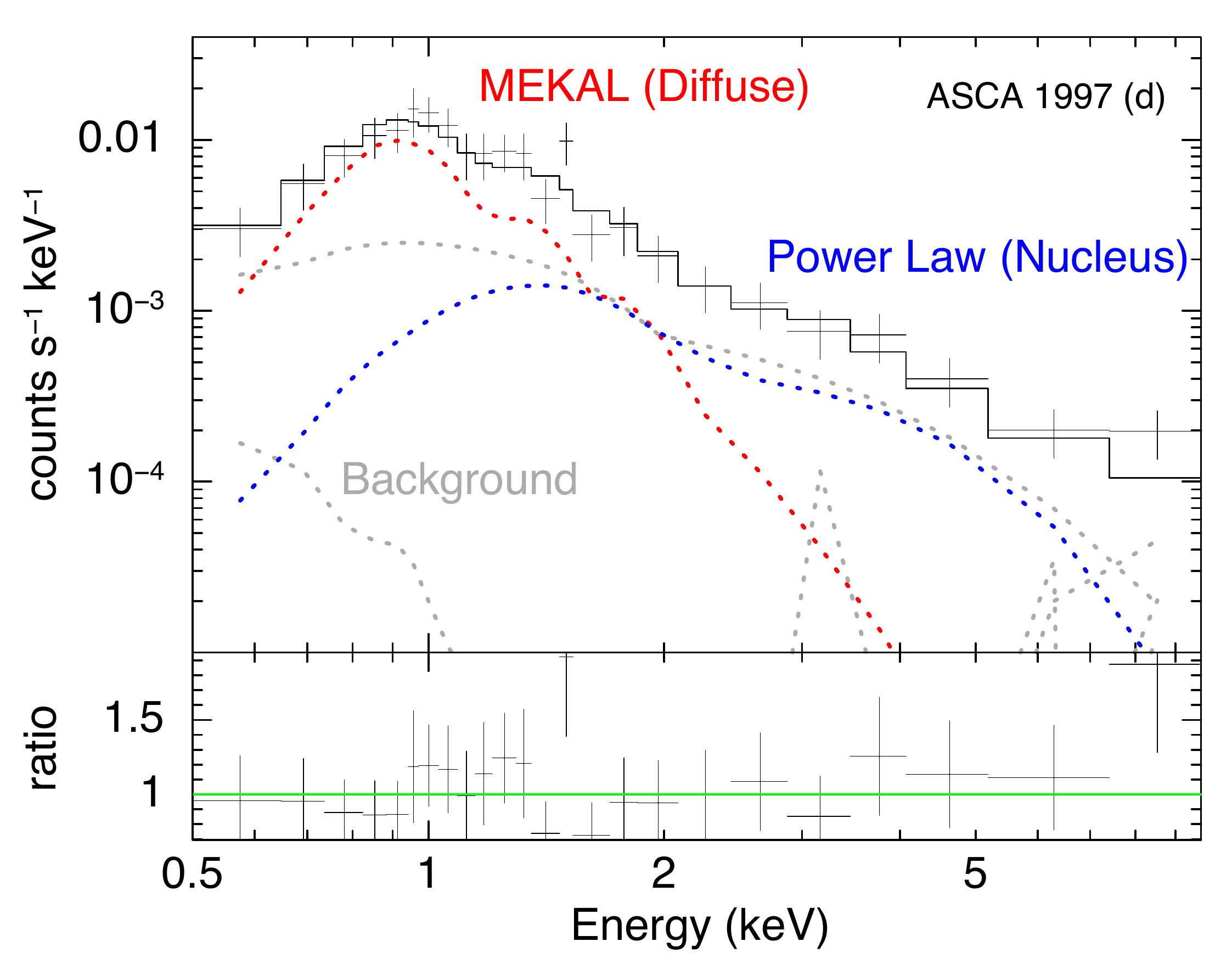}}
\caption{X-ray spectra including model fits and residuals, for four
  X-ray observations of He 2--10. Each component of the model fitted
  to the data are shown as dotted lines: the diffuse VMEKAL component
  (red) and nuclear power law (blue). For the \chandra\ observation
  (a) the background was subtracted before spectral fitting, while for
  the \xmm\ (b), (c) and \asca\ (d) observations, the background
  spectra are fitted by a model(shown by the gray lines)
  simultaneously with fitting of the observed source spectrum.  There
  is a strong and significant detection of the hard nuclear power law
  component in the 2001 \chandra\ (a) observation (clearly visible as
  a point source in Figure~\ref{fig:area}). The hard component is
  significantly weaker in the {\em XMM} observations (b), (c)
  indicating variability by approximately an order of
  magnitude. The hard component is also detected, at lower
  significance, in the 1997 \asca\ spectrum (d).
\label{fig:spectra}}
\end{figure*}

As discussed below, the fluxes of the nuclear power law component in
the {\em XMM} observations are significantly smaller than that
observed for {\em Chandra}.  To most accurately extract the weak hard
X-ray signal from the significant background, we modeled the data by
simultaneously fitting the source and background spectra, using the
model given in Equation~\ref{eqn:xmm}. We account for the differences
in area between the source and background regions by setting the {\tt
  AREASCAL} parameter on the background spectrum. The spectral
parameters of the instrumental and sky backgrounds were fixed to be
equal for the source and background spectra. The source component is
also included in the background spectrum, but multiplied by a factor
$5\times10^{-3}$ to approximately model the flux scattered outside the
source region into the background region. (The scaling factor accounts
for both the energy encircled fraction and the relative area of the
source and background regions; the ultimate fit parameters are
insensitive to the precise value of this factor.)

Because the soft (VMEKAL) emission from the source is diffuse (with a
diameter of $\approx$5\arcsec or 200 pc), and thus a light-crossing
time much longer than the separation in time between observations, it
should not be observed to vary in our data. Therefore, to maximize the
statistical power of our modeling, we fitted the 2004 and 2011
{\em XMM} spectra simultaneously, tying the temperatures and
normalizations of the VMEKAL and APEC components between the two data
sets. We allowed the the nuclear power-law normalization to float, along
with the instrumental background parameters (the particle background
in the detectors should not be perfectly constant for the duration of
the mission). We thus performed a simultaneous fit to four spectra:
source and background spectra from each of the 2004 and 2011.

\begin{figure}
\epsscale{1.1}
\plotone{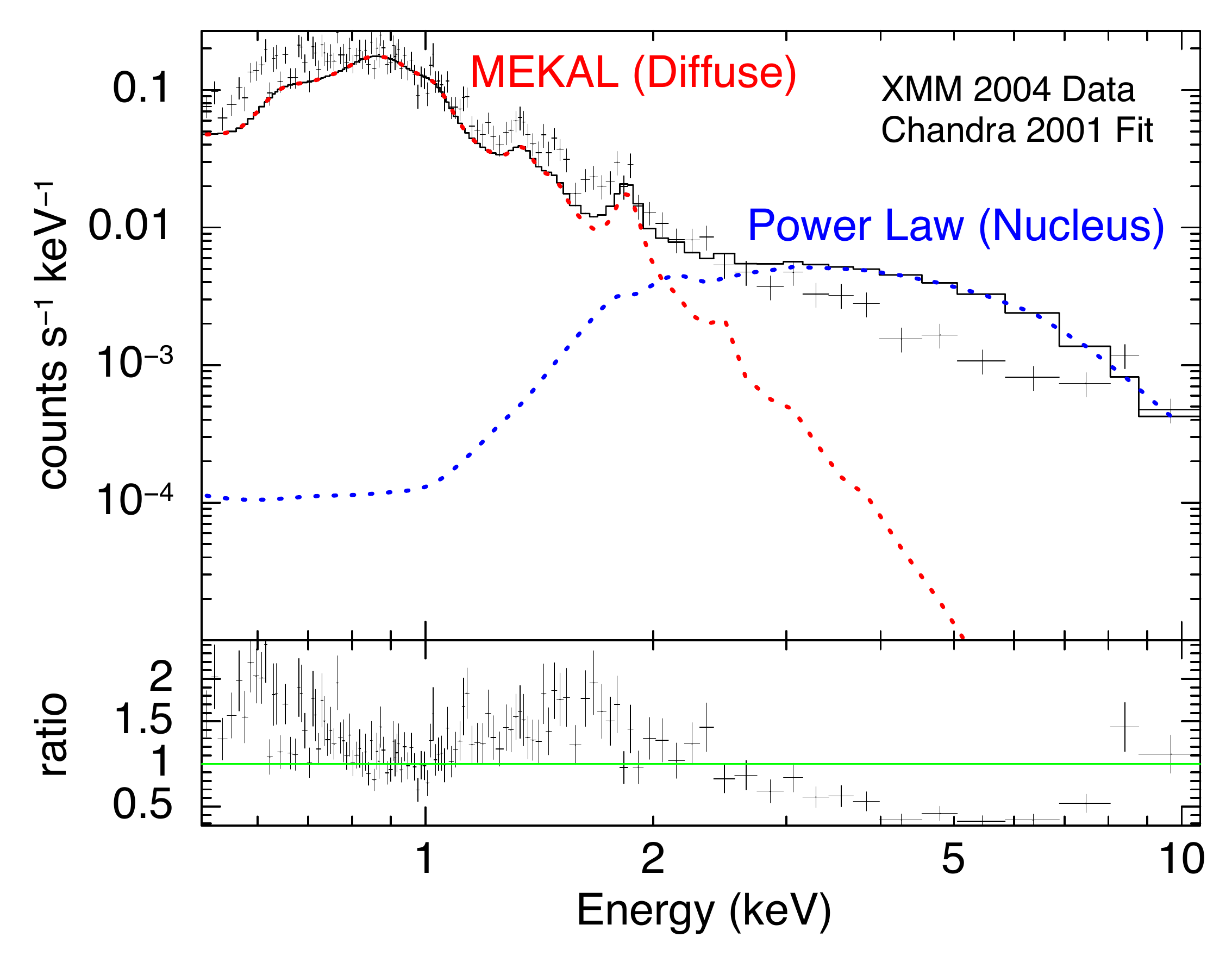}
\caption{Spectrum of the 2004 {\em XMM} observation, fitted with
  source spectrum from the 2001 {\em Chandra} observation. This fit
  includes no subtraction or modeling of the {\em XMM}
  background. This large excess of the model over the data at high
  energies clearly illustrates the dramatic decrease in the flux of
  the hard nuclear component between 2001 and 2004, independent of the
  methods used to account for the {\em XMM} background.
\label{fig:compare}}
\end{figure}

The results of the spectral fitting are shown in
Figure~\ref{fig:spectra} and listed in Table~\ref{table:params}. We
observe a clear decrease in the flux of the hard nuclear component
between the 2001 {\em Chandra} and 2004 {\em XMM} observations. This
is demonstrated in Figure~\ref{fig:compare}, in which we show the 2004
{\em XMM} pn spectrum fitted with the 2001 {\em Chandra} best-fit model, with
no model for the {\em XMM} background included. This shows that 
the hard source flux has dropped dramatically from the {\em Chandra}
level, even before accounting for the {\em XMM} background. We observe
a further, less significant decrease between the 2004 and 2011
\xmm\ observations, while the flux of the diffuse (VMEKAL) component
is consistent with no variation from the {\em Chandra} observation.
We also observe evidence for a decrease in absorption on the nuclear
component, with $N_{\rm H,Nuclear}$ consistent with zero. The best-fit
$kT$ of the VMEKAL component for {\em XMM} is close but not fully
statistically consistent with the {\em Chandra} data
($0.58^{+0.02}_{-0.03}$ keV compared to $0.65\pm0.03$ keV). Fixing
this temperature to the {\em Chandra} value has a negligible effect on
the flux of the diffuse component, but decreases the flux of the hard
nuclear component by $\approx$15\%. This results in an even larger
observed drop in flux compared to the {\em Chandra} data; in the
following discussion we will conservatively consider the smaller
change in flux obtained when the VMEKAL $kT$ allowed to float for
\xmm. The flux of the APEC component, representing the soft diffuse
CXB, corresponds to a 0.5--2 keV surface brightness of
$(2.9\pm0.5)\times10^{-12}$ erg cm$^{-2}$ s$^{-1}$ deg$^{-2}$, similar
to the soft background intensity obtained in the {\em Chandra} Deep
Fields \citep{hick06a}. The overall implications of the spectral
fitting results are discussed in \S~\ref{sec:results}.

One potential uncertainty in our spectral analysis arises from the
implicit assumption that the instrumental and sky background in the
source region has the same surface brightness to the emission in the
background region. It is possible that fluctuations in the background
could cause this assumption to be invalid, leading to over- or
under-subtraction of the background, particularly at energies $>$2 keV
where the background dominates the signal. To directly check the level
of possible fluctations, we extracted the 2--10 keV counts in 80
circular regions of 36\arcsec\ radius (equal to the source region)
surrounding the source region. We avoid chip gaps and obvious bright
sources, noting that there is no bright point source detected within
36\arcsec\ in the 2--8 keV {\em Chandra} image
(Figure~\ref{fig:area}). The 2--10 keV counts in these apertures are
approximately normally distributed, with mean (dispersion) of 239 (19)
and 138 (15) counts for 2004 and 2011, respectively.  We conclude that
our modeling of the background in the source region is dominated by
statistical error rather than any systematic uncertainty due to
background fluctuations.

\subsection{\asca}

As a measurement of the baseline level of flux prior to the {\em
  Chandra} observations, we also utilize observations of He 2--10 from
\asca\ in 1997. Due to \asca's poor angular resolution and sensitivity
relative to {\em Chandra} and {\em XMM}, it provides relatively weak
constraints on the source flux, particularly in the hard nuclear
component. We will therefore focus our conclusions primarily on the
{\em Chandra} and {\em XMM} data, but will utilize the {\em ASCA} data
here as a useful check on our conclusions.

He 2--10 was observed by \asca\ for a total of 39.8 ks on 30 November
1997. We used Xselect v2.4b to remove time intervals of high
background for a net exposure of 22.4 ks, and to extract source and
background spectra from the cleaned event files. We extracted a
spectrum in the energy range between 0.5--10.5 keV, using a source
extraction region of 1.45 arcmin in radius. This source region was
chosen to contain as much of the source flux as possible (the energy
encircled fraction is 50\%) while minimizing the contribution from
background. We extracted a background spectrum from a rectangular
annulus of area $\approx$5 times that of the source region, located
around the source position but excluding the source region. The RMF
and ARF response files were generated for each chip using the commands
{\tt sisrmg} and {\tt ascaarf}, respectively. Furthermore, the spectra
from the two chips SIS0 and SIS1 were added using the HEASOFT command
{\tt mathpha}, while the response files were added with area dependent
weights using the commands {\tt addarf} and {\tt addrmf}. As discussed
in \S~\ref{sec:xmm}, the source ARF includes a correction for the energy
encircled fraction, while the background ARF does not.

As with {\em XMM}, we modeled the \asca\ data by fitting the source
and background spectra simultaneously, modeling the scattered flux in
the background region assuming an energy-encircled fraction of
0.5. The instrumental background was modeled with a power law plus
five Gaussian emission lines; three were included in the model as
narrow fluorescence lines stemming from the device itself at 6.5 keV,
7.5 keV, and 8.2 keV, which are the Fe and Ni K-$\alpha$ lines and the
Ni K-$\beta$
line.\footnote{https://heasarc.gsfc.nasa.gov/docs/asca/newsletters/sis\_back2.html}
Another Gaussian at 3.3 keV of unknown origin was introduced to fit a
feature of the background spectrum, while the final Gaussian was a
broad line at 11 keV that modeled the internal background above 7 keV
to account for the steepness of the power law.

We utilized the same source and sky background models as for
\xmm\ (Equations \ref{eqn:source} and \ref{eqn:skybg}). Because they
cannot be well-constrained owing to the poor photon statistics, the
surface brightness of the APEC component and the VMEKAL temperature
are fixed to the values obtained with \xmm, and the photon index of
the power law is again fixed to $\Gamma=1.8$. (Allowing the VMEKAL
temperature to float yields a significantly higher $kT\approx0.9$,
inconsistent with the \xmm\ and \chandra\ results, but has a
negligible effect on the total fluxes of the diffuse and nuclear
components.) The results of the \asca\ spectral fitting are shown in
Figure~\ref{fig:spectra} and listed in Table~\ref{table:params}. We
obtain a significant detection of both the diffuse and nuclear
spectral components, although with significantly larger uncertainties
than in the \chandra\ and \xmm\ data. In contrast to the
\xmm\ observations, the fit yields significant nuclear absorption
consistent with the {\em Chandra} value, although the precise value of
$N_{\rm H}$ is poorly constrained.

\begin{deluxetable*}{cccccccc}
\tabletypesize{\scriptsize}
\tablecolumns{10}
\tablecaption{Spectral fitting results}
\tablehead{
& \multicolumn{2}{c}{VMEKAL (Diffuse)} &  \multicolumn{4}{c}{Power Law (Nuclear)} & \\
	\colhead{Observation}  & \colhead{$kT$} & \colhead{$F_{\rm 0.5-3\; keV}$}  & \colhead{$N_{\rm H,Nuclear}$} & \colhead{$\Gamma$} & \colhead{$F_{\rm 2-10\;keV}$}   & \colhead{$L_{\rm 2-10\;keV}$\tablenotemark{a}} & \colhead{$\chi^2_\nu$ (d.o.f.)} \\
& (keV) & ($10^{-13}$ erg cm$^{-2}$ s$^{-1}$) & ($10^{22}$ cm$^{-2}$) & & ($10^{-13}$ erg cm$^{-2}$ s$^{-1}$) & $ (10^{39}$ erg s$^{-1}$)}
\startdata
\emph{Chandra}\tablenotemark{b}  (2001) & $0.65\pm0.03$        & $2.06\pm0.11$         & $4.61^{+1.67}_{-1.26}$    & [1.8] &  $3.28^{+0.73}_{-0.64}$      &  $3.18^{+0.71}_{-0.62}$ & 1.61\phn (46)\\ [1ex]
\emph{XMM}\tablenotemark{c} 	(2004) & $0.58^{+0.02}_{-0.03}$ & $1.90^{+0.08}_{-0.05}$  & $<0.05$  & [1.8]        & $0.68^{+0.08}_{-0.07}$  & $0.66^{+0.08}_{-0.07}$ &  1.20 (661)\\ [1ex]
\emph{XMM}\tablenotemark{c} 	(2011) & $0.58^{+0.02}_{-0.03}$ & $1.90^{+0.08}_{-0.05}$  & $<0.05$  & [1.8]        & $0.44^{+0.09}_{-0.10}$  & $0.43\pm0.09$ &  1.20 (661)\\
\emph{ASCA} 	(1997) & [0.65]                 & $2.89^{+0.89}_{-1.02}$  & $0.30^{+3.83}_{-0.30}$    & [1.8]  & $1.79^{+1.66}_{-0.87}$  & $1.73^{+1.61}_{-0.84}$ & 0.80\phn (70)
\enddata

\tablecomments{Best-fit spectral parameters obtained from modeling of the
  four X-ray spectra. The  (unabsorbed) 2--10 keV fluxes were
  calculated based on the nuclear (power law) component, while the
   (observed) 0.5--3 keV fluxes correspond to the diffuse (VMEKAL)
  component. Both components are modified by Galactic absorption with column density fixed at $N_{\rm H,Gal} = 9\times10^{20}$ cm$^{-2}$, and the VMEKAL component is absorbed by an additional component with column density fixed at $N_{\rm
  H,Diffuse} = 9.7\times10^{20}$ cm$^{-2}$. Parameters in the table that are fixed in the fits are identified with brackets. Uncertainties represent 90\% confidence intervals.}
\tablenotetext{a}{~Luminosity values were calculated assuming a distance of 9 Mpc to He 2--10.}
\tablenotetext{b}{Parameters in the {\em Chandra} fits 
for the nuclear and diffuse components are determined from the fit to
the nuclear and extended source regions, respectively, as described in \S~\ref{sec:chandra}.}
\tablenotetext{c}{The spectra for the two {\em XMM} observations are fitted simultaneously, with parameters for the diffuse (MEKAL) component tied between the two observations, as described in \S~\ref{sec:xmm}. \label{table:params}}
\end{deluxetable*}

\section{Results and discussion}

\label{sec:results}

\begin{figure}[b]
\epsscale{1.1}
\plotone{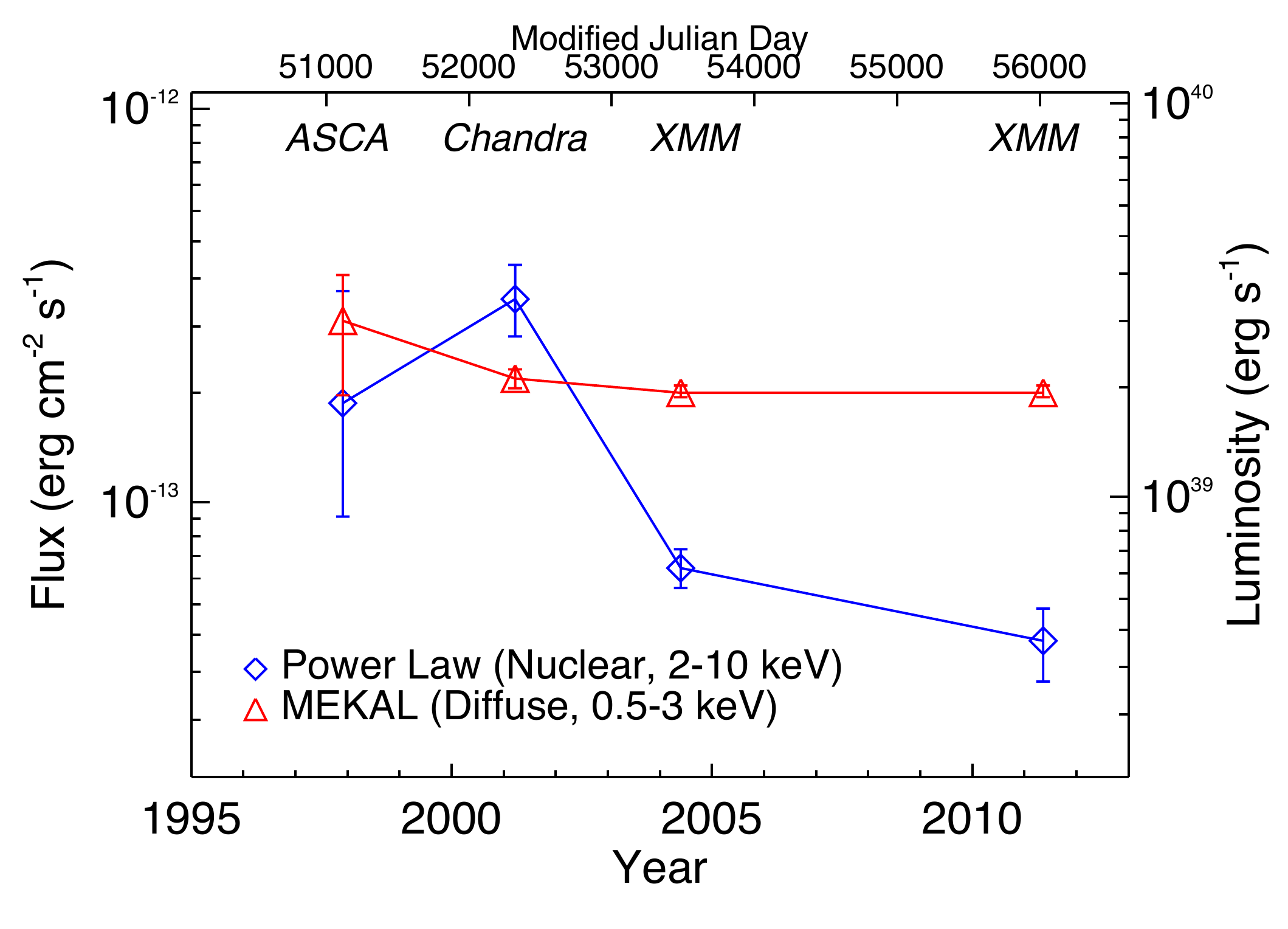}
\caption{The fourteen year light curve of He 2--10 with 90\%
  confidence errors shown. The hard nuclear flux (blue) declines
  between the 2001 \chandra\ and 2011 \xmm\ observations by nearly an
  order of magnitude. In contrast, the soft diffuse flux (red),
  remains approximately constant between the four observations. }
\label{fig:crv}
\end{figure}

The long-term X-ray light curve of He 2--10, showing variations in the
VMEKAL and power law components, are shown in Figure ~\ref{fig:crv}.
The diffuse component shows no significant variability over the four
observations, as expected for emission from a large-scale diffuse
plasma. In contrast, it is immediately clear that there is significant
variability in the hard power law component. The luminosity of the
nuclear source decreased significantly between the 2001 {\em Chandra}
and 2004 {\em XMM} observations, and by approximately an order of magnitude between
2001 and the 2011 \xmm\ observation. (We note that our overall
conclusions are unchanged if we use observed hard fluxes, which differ
by roughly 30\% from the values shown in the Table 2 for the
\chandra\ and \asca\ observation and remain constant for \xmm\ due to
the decreased levels of obscuration.)

The variation in the light curve of hard spectral component He 2--10
over approximately an order of magnitude in $L_X$ confirms that this
emission is the result of a single object, rather than several
separate sources. This allows us to perform comparisons to other
individual astrophysical sources. One class of object that can have
similar X-ray and luminosities and amplitudes of variability is
SNe (see \citealt{dwar12} for a compilation of published SN
X-ray light curves).  However, a SN interpretation for the nuclear
source is inconsistent with the radio properties. The radio flux of He
2--10 has been measured at 5 GHz with the VLA in 1994 ($0.89\pm0.18$
mJy; \citealt{kobu99}) and 2004 ($0.86\pm0.02$ mJy;
\citealt{rein11he2}), implying no significant change in the radio flux
with time.  Long Baseline Array measurements in 2011 at 1.4 GHz yield
a flux of $0.98\pm0.21$ mJy in a compact source on $\approx$1 pc
spatial scales; assuming a typical radio spectral index, this implies
that roughly half of the observed 5 GHz flux comes from the central
compact source \citep{rein12he2}.  High Sensitivity Array observations
in 2005 did not detect the compact source on extremely small
($\sim$0.1 pc) spatial scales \citep{ulv07wrradio}, placing a lower
limit on its spatial extent. This rules out out the presence of a
single very young SN, however the total radio luminosity of the source
would imply that any single SN must be at most decades old
\citep{fene10snm82}.

With these constraints in mind, we test whether the nuclear source in
He 2--10 is consistent with a (relatively) evolved SN explosion, by
comparing its (unabsorbed) soft X-ray and 5 GHz radio light curves to
a sample of 7 SNe that have both radio and X-ray measurements in the
compilations of \citet{dwar12} and \citet{weil02}. Assuming that the
nuclear source in He 2--10 has a relatively constant 5 GHz flux
between 1994 and 2004, and conservatively assigning all the 5 GHz flux
observed with the VLA ($\approx$0.9 mJy) to the central compact
component, we find that the ratio of X-ray (0.5--2 keV) to radio ($\nu
F_\nu$ at 5 GHz) for He 2--10 is $\sim$$5\times10^{3}$. This is more
than an order of magnitude larger than the typical X-ray to radio flux
ratio for X-ray-detected SNe at ages $>$1 year, and 2.5 times larger
than the most extreme observed values, in SNe 1980K and
1970G. Further, the detection of the nuclear hard X-ray component in
the \asca\ observation indicates that the X-ray light curve rises or
remains constant, and then declines sharply over a few years. This
behavior is unusual for SNe; of the eight SN X-ray light curves in the
compilation of \citet{dwar12} that extend beyond ten years, none shows
a similar sudden, rapid decline on these time scales.  We also note
that the X-ray to radio flux ratio is consistently at least 2--3
orders of magnitude {\em smaller} than the typical ratio of X-ray to
compact radio flux for ULXs, for which the compact sources tend to be
weak or undetected in the radio \citep[e.g.,][]{midd13ulx, wolt14ulx}.
Given the observed X-ray light curve and the ratio of X-ray and radio
luminosities, we conclude that the observations do not favor a SN or
ULX origin for the nuclear source.
 
In contrast, the significant variability of the hard nuclear X-ray
source in He 2--10 is consistent with its identification as an
accreting massive BH, in comparison to the X-ray variability of known
low-mass AGN. The Sdm spiral galaxy NGC 4395, at a distance of only 4
Mpc, contains a BH of mass $3.6 \times 10^5 \msun$, whose hard
component, as is shown in \citet{king13}, can vary by a factor of 5 on
a timescale of just one day. The nearby edge-on Seyfert 2 galaxy NGC
4945, with a BH mass obtained through observations of its H$_2$0
megamaser of $\approx 10^6 \msun$, shows intrinsic variability
(measured at $>$8 keV by {\em RXTE} and {\em Swift}/BAT) of at least
an order of magnitude on timescales of days to weeks
\citep[e.g.,][]{muel04ngc4945,mari12ngc4945}. The AGN in the nearby
Seyfert 1 galaxy NGC 4051 ($M_{\rm BH} = 1.73 \times 10^6 \msun$;
\citealt{denn09}) has been observed to vary in X-ray luminosity by
more than an order of magnitude over $\sim$\,year timescales
\citep{uttl99}. This limited survey confirms that X-ray variability
over a large dynamic range on timescales of years is not uncommon
among relatively low-mass AGNs. Therefore, the decreased observed
luminosity after the \chandra\ observations could either be due to
short timescale fluctuations occurring precisely at the time of the
observation, as in NGC 4945, or part of a general trend of long
timescale variability, similar to objects like NGC 4051.

\acknowledgements{
Support for A.E.R.\ was provided by NASA through the Einstein Fellowship Program, grant PF1-120086, and the Hubble Fellowship Program. R.C.H.\ acknowledges support from the Dartmouth College Class of 1962 Faculty Fellowship. G.R.S.\ acknowledges support from an NSERC Discovery Grant. This research has made use of data, software and/or web tools obtained from NASA's High Energy Astrophysics Science Archive Research Center (HEASARC), a service of Goddard Space Flight Center and the Smithsonian Astrophysical Observatory.}


\begin{thebibliography}{}
\expandafter\ifx\csname natexlab\endcsname\relax\def\natexlab#1{#1}\fi

\bibitem[{{Allen} {et~al.}(1976){Allen}, {Wright}, \& {Goss}}]{alle76}
{Allen}, D.~A., {Wright}, A.~E., \& {Goss}, W.~M. 1976, \mnras, 177, 91

\bibitem[{{Arnaud}(1996)}]{arna96xsp}
{Arnaud}, K.~A. 1996, in Astronomical Society of the Pacific Conference Series,
  Vol. 101, Astronomical Data Analysis Software and Systems V, ed. G.~H.
  {Jacoby} \& J.~{Barnes}, 17

\bibitem[{{Carter} \& {Read}(2007)}]{cart07xmm}
{Carter}, J.~A., \& {Read}, A.~M. 2007, \aap, 464, 1155

\bibitem[{{Denney} {et~al.}(2009){Denney}, {Watson}, {Peterson}, {Pogge},
  {Atlee}, {Bentz}, {Bird}, {Brokofsky}, {Comins}, {Dietrich}, {Doroshenko},
  {Eastman}, {Efimov}, {Gaskell}, {Hedrick}, {Klimanov}, {Klimek}, {Kruse},
  {Lamb}, {Leighly}, {Minezaki}, {Nazarov}, {Petersen}, {Peterson},
  {Poindexter}, {Schlesinger}, {Sakata}, {Sergeev}, {Tobin}, {Unterborn},
  {Vestergaard}, {Watkins}, \& {Yoshii}}]{denn09}
{Denney}, K.~D., {Watson}, L.~C., {Peterson}, B.~M., {et~al.} 2009, \apj, 702,
  1353

\bibitem[{{Dewangan} {et~al.}(2008){Dewangan}, {Mathur}, {Griffiths}, \&
  {Rao}}]{dewa08}
{Dewangan}, G.~C., {Mathur}, S., {Griffiths}, R.~E., \& {Rao}, A.~R. 2008,
  \apj, 689, 762

\bibitem[{{Dwarkadas} \& {Gruszko}(2012)}]{dwar12}
{Dwarkadas}, V.~V., \& {Gruszko}, J. 2012, \mnras, 419, 1515

\bibitem[{{Fenech} {et~al.}(2010){Fenech}, {Beswick}, {Muxlow}, {Pedlar}, \&
  {Argo}}]{fene10snm82}
{Fenech}, D., {Beswick}, R., {Muxlow}, T.~W.~B., {Pedlar}, A., \& {Argo}, M.~K.
  2010, \mnras, 408, 607

\bibitem[{{Hickox} \& {Markevitch}(2006)}]{hick06a}
{Hickox}, R.~C., \& {Markevitch}, M. 2006, \apj, 645, 95

\bibitem[{{Hickox} \& {Markevitch}(2007)}]{hick07b}
---. 2007, \apjl, 661, L117

\bibitem[{{Johnson} \& {Bromm}(2007)}]{john07firstbh}
{Johnson}, J.~L., \& {Bromm}, V. 2007, \mnras, 374, 1557

\bibitem[{{Johnson} \& {Kobulnicky}(2003)}]{john03he2}
{Johnson}, K.~E., \& {Kobulnicky}, H.~A. 2003, \apj, 597, 923

\bibitem[{{King} {et~al.}(2013){King}, {Miller}, {Reynolds}, {G\"{u}letkin},
  {Gallo}, \& {Dipankar}}]{king13}
{King}, A.~K., {Miller}, J.~M., {Reynolds}, M.~T., {et~al.} 2013, \apjl, 774,
  L25

\bibitem[{{Kobulnicky} \& {Johnson}(1999)}]{kobu99}
{Kobulnicky}, H.~A., \& {Johnson}, K.~E. 1999, \apj, 527, 154

\bibitem[{{Kobulnicky} \& {Martin}(2010)}]{kobu10he2}
{Kobulnicky}, H.~A., \& {Martin}, C.~L. 2010, \apj, 718, 724

\bibitem[{{Lemons} {et~al.}(2015){Lemons}, {Reines}, {Plotkin}, {Gallo}, \&
  {Greene}}]{lemo15xdwarf}
{Lemons}, S., {Reines}, A., {Plotkin}, R., {Gallo}, E., \& {Greene}, J. 2015,
  \apj\ in press (arXiv:1502.06958), arXiv:1502.06958

\bibitem[{{Liedahl} {et~al.}(1995){Liedahl}, {Osterheld}, \&
  {Goldstein}}]{lied95mekal}
{Liedahl}, D.~A., {Osterheld}, A.~L., \& {Goldstein}, W.~H. 1995, \apjl, 438,
  L115

\bibitem[{{Marinucci} {et~al.}(2012){Marinucci}, {Risaliti}, {Wang}, {Nardini},
  {Elvis}, {Fabbiano}, {Bianchi}, \& {Matt}}]{mari12ngc4945}
{Marinucci}, A., {Risaliti}, G., {Wang}, J., {et~al.} 2012, \mnras, 423, L6

\bibitem[{{Merloni} {et~al.}(2003){Merloni}, {Heinz}, \& {di
  Matteo}}]{merl03bhplane}
{Merloni}, A., {Heinz}, S., \& {di Matteo}, T. 2003, \mnras, 345, 1057

\bibitem[{{Mewe} {et~al.}(1985){Mewe}, {Gronenschild}, \& {van den
  Oord}}]{mewe85mekal}
{Mewe}, R., {Gronenschild}, E.~H.~B.~M., \& {van den Oord}, G.~H.~J. 1985,
  \aaps, 62, 197

\bibitem[{{Middleton} {et~al.}(2013){Middleton}, {Miller-Jones}, {Markoff},
  {Fender}, {Henze}, {Hurley-Walker}, {Scaife}, {Roberts}, {Walton},
  {Carpenter}, {Macquart}, {Bower}, {Gurwell}, {Pietsch}, {Haberl}, {Harris},
  {Daniel}, {Miah}, {Done}, {Morgan}, {Dickinson}, {Charles}, {Burwitz}, {Della
  Valle}, {Freyberg}, {Greiner}, {Hernanz}, {Hartmann}, {Hatzidimitriou},
  {Riffeser}, {Sala}, {Seitz}, {Reig}, {Rau}, {Orio}, {Titterington}, \&
  {Grainge}}]{midd13ulx}
{Middleton}, M.~J., {Miller-Jones}, J.~C.~A., {Markoff}, S., {et~al.} 2013,
  \nat, 493, 187

\bibitem[{{Mueller} {et~al.}(2004){Mueller}, {Madejski}, {Done}, \&
  {Zycki}}]{muel04ngc4945}
{Mueller}, M., {Madejski}, G., {Done}, C., \& {Zycki}, P. 2004, in American
  Institute of Physics Conference Series, Vol. 714, X-ray Timing 2003: Rossi
  and Beyond, ed. P.~{Kaaret}, F.~K. {Lamb}, \& J.~H. {Swank}, 190--193

\bibitem[{{Reines} \& {Deller}(2012)}]{rein12he2}
{Reines}, A.~E., \& {Deller}, A.~T. 2012, \apjl, 750, L24

\bibitem[{{Reines} {et~al.}(2013){Reines}, {Greene}, \& {Geha}}]{rein13dwarf}
{Reines}, A.~E., {Greene}, J.~E., \& {Geha}, M. 2013, \apj, 775, 116

\bibitem[{{Reines} {et~al.}(2014){Reines}, {Plotkin}, {Russell}, {Mezcua},
  {Condon}, {Sivakoff}, \& {Johnson}}]{rein14mrk709}
{Reines}, A.~E., {Plotkin}, R.~M., {Russell}, T.~D., {et~al.} 2014, \apjl, 787,
  L30

\bibitem[{{Reines} {et~al.}(2011){Reines}, {Sivakoff}, {Johnson}, \&
  {Brogan}}]{rein11he2}
{Reines}, A.~E., {Sivakoff}, G.~R., {Johnson}, K.~E., \& {Brogan}, C.~L. 2011,
  \nat, 470, 66

\bibitem[{{Roberts}(2007)}]{robe07ulx}
{Roberts}, T.~P. 2007, \apss, 311, 203

\bibitem[{{Secrest} {et~al.}(2015){Secrest}, {Satyapal}, {Gliozzi}, {Rothberg},
  {Ellison}, {Mowry}, {Rosenberg}, {Fischer}, \& {Schmitt}}]{secr15xdwarf}
{Secrest}, N.~J., {Satyapal}, S., {Gliozzi}, M., {et~al.} 2015, \apj, 798, 38

\bibitem[{{Smith} {et~al.}(2001){Smith}, {Brickhouse}, {Liedahl}, \&
  {Raymond}}]{smit01}
{Smith}, R.~K., {Brickhouse}, N.~S., {Liedahl}, D.~A., \& {Raymond}, J.~C.
  2001, \apjl, 556, L91

\bibitem[{{Tozzi} {et~al.}(2006){Tozzi}, {Gilli}, {Mainieri}, {Norman},
  {Risaliti}, {Rosati}, {Bergeron}, {Borgani}, {Giacconi}, {Hasinger},
  {Nonino}, {Streblyanska}, {Szokoly}, {Wang}, \& {Zheng}}]{tozz06}
{Tozzi}, P., {Gilli}, R., {Mainieri}, V., {et~al.} 2006, \aap, 451, 457

\bibitem[{{Ulvestad} {et~al.}(2007){Ulvestad}, {Johnson}, \&
  {Neff}}]{ulv07wrradio}
{Ulvestad}, J.~S., {Johnson}, K.~E., \& {Neff}, S.~G. 2007, \aj, 133, 1868

\bibitem[{{Uttley} {et~al.}(1999){Uttley}, {M\textsuperscript{c}Hardy},
  {Papadakis}, {Guainazzi}, \& {Fruscione}}]{uttl99}
{Uttley}, P., {M\textsuperscript{c}Hardy}, I.~M., {Papadakis}, I.~E.,
  {Guainazzi}, M., \& {Fruscione}, A. 1999, \mnras, 307, L6

\bibitem[{{Volonteri}(2010)}]{volo10bhform}
{Volonteri}, M. 2010, \aapr, 18, 279

\bibitem[{{Weiler} {et~al.}(2002){Weiler}, {Panagia}, {Montes}, \&
  {Sramek}}]{weil02}
{Weiler}, K.~W., {Panagia}, N., {Montes}, M.~J., \& {Sramek}, R.~A. 2002,
  \araa, 40, 387

\bibitem[{{Wilms} {et~al.}(2000){Wilms}, {Allen}, \& {McCray}}]{wilm00tbabs}
{Wilms}, J., {Allen}, A., \& {McCray}, R. 2000, \apj, 542, 914

\bibitem[{{Wolter} {et~al.}(2014){Wolter}, {Rushton}, {Mezcua}, {Cseh},
  {Pintore}, {Prandoni}, {Paragi}, \& {Zampieri}}]{wolt14ulx}
{Wolter}, A., {Rushton}, A.~P., {Mezcua}, M., {et~al.} 2014, PoS in press
  (arXiv:1412.5643)


\end{thebibliography}
\end{document}